\documentclass[11pt]{article}

\usepackage[a4paper,margin=1in]{geometry}
\usepackage{amsmath,amssymb,amsfonts,amsthm}
\usepackage{mathrsfs}
\usepackage[T1]{fontenc}
\usepackage[utf8]{inputenc}
\usepackage{textcomp}
\usepackage[expansion=false]{microtype}
\usepackage{xcolor}
\usepackage{listings}
\usepackage{verbatim}
\usepackage{authblk}
\usepackage[square,authoryear,sort&compress]{natbib}
\usepackage[hidelinks,
            colorlinks=true,
            linkcolor=black,
            citecolor=black,
            urlcolor=black,
            filecolor=black,
            anchorcolor=black]{hyperref}

\lstdefinestyle{coq}{
  basicstyle    = \ttfamily\small,
  keepspaces    = true,
  columns       = flexible,
  showstringspaces = false,
  breaklines    = true,
  xleftmargin   = 1.2em,
  aboveskip     = 0.8em,
  belowskip     = 0.8em,
  frame         = none
}
\bibpunct{[}{]}{,}{a}{,}{,}

\renewcommand{\Affilfont}{\normalsize}
\makeatletter
\renewcommand\AB@affilsepx{\protect\\ \protect\Affilfont}
\makeatother

\title{Four Paradoxes and a Proof Assistant: \\
       Burali-Forti, Diaconescu, Reynolds, and Hurkens \\
       in the \texttt{coq-paradoxes} library}

\author[1]{Bernardo Alonso}
\affil[1]{Department of Philosophy, Universidade Federal de Mato Grosso, Brazil}
\affil[ ]{\texttt{bernardo.alonso@ufmt.br}}

\date{\today}

\begin{document}

\maketitle

\begin{abstract}
\noindent This article reads the four paradoxes mechanised in the \texttt{coq-paradoxes} package \citep{barras2019paradoxes}, namely the Burali-Forti paradox in system U, the Diaconescu paradox that the axiom of choice entails excluded middle, the Reynolds paradox that System F has no set-theoretic model, and the Hurkens paradox in system U as adapted by \citet{geuvers2001inconsistency} for impredicative Set. The package collects four pieces of mechanised mathematics that, taken together, draw the boundary of the Calculus of Inductive Constructions from the outside: each file formalises a derivation of False in a system close to CIC, and each shows where the kernel of Rocq has been designed to refuse to compile the construction. The article walks through the shared machinery of well-foundedness in \emph{Logics.v}, reads the Burali-Forti construction in \emph{BuraliForti.v} against \citeauthor{coquand1986analysis}'s \citeyearpar{coquand1986analysis} analysis of Girard's paradox, sets out the Diaconescu argument as \citet{werner1997sets} implements it in \emph{diaconescu.v}, reconstructs the Reynolds argument in \emph{Reynolds.v} via the preinitial PHI-algebra and Lawvere's fixed-point theorem, and follows Geuvers's adaptation of Hurkens in \emph{Hurkens\_Set.v}. The four together establish three boundary conditions on the kernel of Rocq: the placement of impredicativity, the restriction of large elimination, and the discipline of universe constraints. The article argues that the package is best read not as a collection of curiosities but as a negative specification of what Rocq's kernel had to be designed to refuse, and as evidence that the refusal is being made for the right reasons.
\end{abstract}


%%==============================================================
\section{Introduction}

There is a small library distributed with Rocq, called \emph{coq-paradoxes}, whose contents are at first sight peculiar. The package mechanises four derivations of False under appropriate hypotheses, and ships them as part of the proof assistant's ecosystem \citep{barras2019paradoxes}. The package does not break the soundness of Rocq, because each derivation requires hypotheses that the kernel itself rejects; what the package does is to make the rejection precise. Read carelessly, the library is a piece of formal mathematics that happens to be inconsistent. Read carefully, it is a negative specification of the system in which it sits, written in the language of the system itself, and it tells the reader exactly where the kernel of Rocq has had to draw its lines.\footnote{The package is currently maintained by Hugo Herbelin and was originally contributed by Bruno Barras, Thierry Coquand, Hugo Herbelin, and Benjamin Werner \citep{barras2019paradoxes}. The opam package \texttt{coq-paradoxes 8.10.0} was published on 7 December 2019 and contains eight Coq files. The Rocq-prover package index continues to host it at \url{https://rocq-prover.org/p/coq-paradoxes/latest}, and the source tree is mirrored at \url{https://github.com/rocq-archive/paradoxes} under LGPL 2.1.}

The four paradoxes are not arbitrary. Burali-Forti, in the form due to \citet{coquand1986analysis}, is the canonical impossibility result for a type theory with too much polymorphism: the existence of a universal type of relations forces an ordinal of all ordinals, which a well-founded order cannot accommodate. Diaconescu, in the form due to \citet{werner1997sets}, shows that an apparently innocent choice principle imports the law of excluded middle into a constructive system, and shows it by writing a small program that decides an arbitrary proposition. \citet{reynolds1984polymorphism} shows that polymorphism is not set-theoretic, in the sense that no set-theoretic model can interpret System F faithfully, and the Barras--Werner reformulation of his argument places the impossibility inside Coq itself as the non-existence of an injection from Prop into a single proposition. \citet{hurkens1995simplification}, in the adaptation by \citet{geuvers2001inconsistency} that the present file mechanises, shows that excluded middle cannot be added to impredicative Set in CIC without producing inconsistency. The four results map cleanly onto four design decisions in the kernel of Rocq, and the present article reads them in that order.

The argument of the article is that the coq-paradoxes package is best understood as a proof, by failed example, of the soundness of the kernel. The kernel admits impredicativity at Prop, restricts large elimination, and enforces universe constraints. Each of these decisions can be motivated abstractly. The package motivates them concretely: each file is a construction that would produce False if the decision had been made the other way, and each file is annotated by the precise mechanism in the kernel that rejects it. The reading of the package as negative specification is not new, but the literature has tended to treat the four paradoxes in isolation, each in the context of its own historical development. The present article reads them together, against the source code, and argues that what they share is more revealing than what distinguishes them.

Section~\ref{sec:shared} sets out the shared machinery of well-foundedness in the file \emph{Logics.v}, and explains why the accessibility predicate is the proof engine that the Burali-Forti paradox needs. Section~\ref{sec:burali} reads the Burali-Forti development in \emph{BuraliForti.v}, culminating in the observation that the natural way to instantiate the universal type of relations triggers a Universe Inconsistency in the Rocq kernel, which is exactly the rejection one wants. Section~\ref{sec:diaconescu} reads the Diaconescu development in \emph{diaconescu.v} and identifies the inductive type at the heart of the argument as a piece of programming rather than a piece of metaphysics. Section~\ref{sec:reynolds} reads the Reynolds development in \emph{Reynolds.v}, which is by some distance the most intricate of the four, and which combines a Church-encoded preinitial algebra with a partial-equivalence-relation quotient to recover Cantor's diagonal argument inside Prop. Section~\ref{sec:hurkens} reads the Hurkens-Geuvers development in \emph{Hurkens\_Set.v} and explains why the retract structure between Prop and a small type produces the contradiction. Section~\ref{sec:boundary} concludes by reading the four results together as a specification, drawing the three design decisions out of the codebase.

%%==============================================================
\section{The Shared Machinery: Well-Foundedness in \texttt{Logics.v}}
\label{sec:shared}

The four paradoxes do not all use the same proof technique, but two of them (Burali-Forti and Reynolds) need a piece of infrastructure that is also of independent interest, namely the accessibility predicate and its consequences. The file \emph{Logics.v} in the coq-paradoxes package contains seventy lines of code that set up this infrastructure, and the reading of the rest of the package is easier if one starts here.

The central definition is the accessibility predicate, written in CIC as follows:
\begin{lstlisting}
Inductive ACC (A : Type) (R : A -> A -> Prop) : A -> Prop :=
  ACC_intro :
    forall x : A, (forall y : A, R y x -> ACC A R y) -> ACC A R x.
\end{lstlisting}

The intended reading is that an element \texttt{x} is accessible under a relation \texttt{R} when every \texttt{R}-predecessor of \texttt{x} is itself accessible. The definition is impredicative in the sense that the recursion is allowed to go down arbitrarily deep, but it is well-founded by construction because the only constructor \texttt{ACC\_intro} requires its premise to be supplied at the point of construction. An element is therefore accessible if and only if there is no infinite descending \texttt{R}-chain starting from it, and an element is \emph{not} accessible if and only if it lies on such a chain. The well-foundedness of \texttt{R} is then the universal accessibility of every element:
\begin{lstlisting}
Definition WF (A : Type) (R : A -> A -> Prop) := forall x : A, ACC A R x.
\end{lstlisting}

From this definition, the file proves a small lemma whose statement is innocuous and whose use is crucial. The lemma is \texttt{ACC\_nonreflexive}, and it says that no accessible element is reflexively related to itself:\footnote{It is worth pausing on what \texttt{ACC\_nonreflexive} is doing logically. The accessibility predicate is a positive inductive definition, and the lemma is proved by case analysis on its single constructor. The case analysis gives back the universally quantified premise of the constructor, and that premise applied to the element itself, together with the hypothesis \texttt{R x x}, recurses on the same element. The recursion is well-founded because the accessibility derivation is being consumed at each step. The lemma is, in effect, a Tarski-style fixed-point argument carried out at the level of the proof object, and it is the workhorse that turns ``this relation is well-founded'' into ``this thing cannot relate to itself.''}
\begin{lstlisting}
Lemma ACC_nonreflexive :
  forall (A : Type) (R : A -> A -> Prop) (x : A),
  ACC A R x -> R x x -> False.
\end{lstlisting}

The proof is by induction on the accessibility derivation. The induction hypothesis says that no predecessor of \texttt{x} is reflexively related to itself; the hypothesis \texttt{R x x} says that \texttt{x} is a predecessor of itself; applying the induction hypothesis to \texttt{x} and to the proof \texttt{R x x} gives False. The argument is two lines of tactics. The consequence is the engine of the Burali-Forti paradox.

The remainder of \emph{Logics.v} sets up the inverse-image construction. Given a function \texttt{f : A -> B} and a relation \texttt{R : B -> B -> Prop}, the inverse image is the relation \texttt{Rof x y := R (f x) (f y)} on \texttt{A}. The file proves that \texttt{Rof} inherits well-foundedness from \texttt{R}, in the precise sense that \texttt{WF B R -> WF A Rof}. This is the construction that lets the Burali-Forti paradox transport its well-founded relation through arbitrary morphisms, and it is also the construction that the Reynolds development uses to set up the per quotient. The file is short, but it pays for the next two hundred lines of the package.

%%==============================================================
\section{Burali-Forti in System U: The Ordinal of Ordinals}
\label{sec:burali}

The Burali-Forti paradox, in the form that \citet{coquand1986analysis} analysed and that \emph{BuraliForti.v} mechanises, takes the following shape. Suppose there is a type \texttt{A0} together with an injection \texttt{i0} of relations on arbitrary types into \texttt{A0}, in the sense that \texttt{i0 X1 R1 = i0 X2 R2} implies the existence of a morphism between \texttt{(X1, R1)} and \texttt{(X2, R2)}. Suppose further that the type \texttt{A0} itself lives in a universe high enough to receive its own image under \texttt{i0}, that is, there exists an inhabitant \texttt{Omega := i0 A0 emb} where \texttt{emb} is the embedding relation on \texttt{A0}. From these assumptions, one can derive False.\footnote{System U is the Pure Type System with two universes $\ast$ and $\square$, in which the abstraction $\lambda \alpha{:}\square. \, t$ and the type formation $\Pi\alpha{:}\square.\, A$ are both allowed, and in which $\square : \triangle$ gives access to a third sort. The variant called System U-minus drops the $\triangle$ sort and is the system in which \citet{hurkens1995simplification} and \citet{coquand1986analysis} work. The Rocq kernel does not implement system U; it implements the Calculus of Inductive Constructions, whose universe rules are strictly weaker. The paradoxes are encoded as assumptions on top of CIC: a hypothesis of the form ``there exists a universal type of relations'' together with the appropriate injection produces False.}

The mechanisation makes the hypothesis precise. The package declares:
\begin{lstlisting}
Section Burali_Forti_Paradox.

  Variable A0 : Type.
  Variable i0 : forall X : Type, (X -> X -> Prop) -> A0.
  Hypothesis inj :
    forall (X1 : Type) (R1 : X1 -> X1 -> Prop)
           (X2 : Type) (R2 : X2 -> X2 -> Prop),
    i0 X1 R1 = i0 X2 R2 ->
    exists f : X1 -> X2, morphism X1 R1 X2 R2 f.
\end{lstlisting}

The hypothesis \texttt{inj} is the substantive one. It says that \texttt{i0} identifies relations only when there is a morphism between them, which is the type-theoretic surrogate for injectivity-modulo-isomorphism. The morphism is, by definition, an order-preserving map: \texttt{morphism X1 R1 X2 R2 f := forall x y, R1 x y -> R2 (f x) (f y)}. Under this hypothesis, the development constructs the embedding relation \texttt{emb} on \texttt{A0} and proves it well-founded.

The embedding relation is defined as a record. Two elements \texttt{x} and \texttt{y} of \texttt{A0} are in \texttt{emb} when \texttt{x} is the image of some well-founded \texttt{(X1, R1)} under \texttt{i0}, \texttt{y} is the image of some \texttt{(X2, R2)}, and there is a strict order morphism from \texttt{(X1, R1)} into a strict initial segment of \texttt{(X2, R2)}. The record carries the morphism \texttt{f}, an upper bound \texttt{maj} in \texttt{X2}, and a proof that \texttt{R2 (f z) maj} holds for every \texttt{z} in \texttt{X1}. The reader who has seen the Veblen normal form for ordinals will recognise the construction: \texttt{emb x y} says that the ordinal coded by \texttt{x} is strictly less than the ordinal coded by \texttt{y}.\footnote{The \texttt{Defined} keyword in \citeauthor{coquand1986analysis}'s Burali-Forti development (rather than \texttt{Qed}) matters. A proof closed with \texttt{Defined} produces a term that the kernel can unfold during conversion, whereas \texttt{Qed} produces an opaque term. The choice is forced by the structure of the paradox: subsequent lemmas need to compute through these proofs to extract the embedding witnesses, and an opaque proof would not let them do so. The whole development reads as a piece of computation as much as a piece of logic, which is exactly the point of working in CIC.}

Well-foundedness of \texttt{emb} is proved by a careful induction. The lemma \texttt{ACC\_emb} says that if \texttt{x} is accessible under \texttt{R} in \texttt{X}, and \texttt{f : Y -> X} is a morphism from \texttt{(Y, S)} to \texttt{(X, R)} such that \texttt{R (f y) x} holds for every \texttt{y}, then \texttt{i0 Y S} is accessible under \texttt{emb}. The proof uses the hypothesis \texttt{inj} to extract a witness morphism whenever two images under \texttt{i0} coincide, and then transports the accessibility derivation along the inverse-image construction from \emph{Logics.v}. The conclusion is the lemma \texttt{WF\_emb : WF A0 emb}.

With well-foundedness in hand, the paradox proceeds. The element \texttt{Omega := i0 A0 emb} is the image of the embedding relation under \texttt{i0}. The development then defines, for each \texttt{a : A0}, the type \texttt{sub a} of elements of \texttt{A0} smaller than \texttt{a} under \texttt{emb}, carrying the witness \texttt{emb\_wit}. The function \texttt{F : A0 -> A0} sends \texttt{a} to \texttt{i0 (sub a) (Rof \_ \_ emb witness)}, that is, to the encoding of the relation ``embed via the witness projection.'' The lemma \texttt{F\_emb\_Omega : emb (F a) Omega} says that \texttt{F a} is embedded in \texttt{Omega} for every \texttt{a}, and the lemma \texttt{F\_morphism} says that \texttt{F} itself is an order morphism. The crucial step is the theorem \texttt{Omega\_refl : emb Omega Omega}: Omega is embedded in itself, because \texttt{F} is a morphism from \texttt{A0} with \texttt{emb} into itself, with upper bound the original \texttt{Omega}.

The contradiction is immediate. The well-founded relation \texttt{emb} has no reflexive elements by \texttt{ACC\_nonreflexive}; the element \texttt{Omega} is reflexive in \texttt{emb} by \texttt{Omega\_refl}; therefore False. The theorem \texttt{Burali\_Forti : False} is proved in four lines.

This is not yet a paradox in CIC. It is a paradox under the hypothesis that \texttt{A0} and \texttt{i0} exist with the stated properties, and the closing comment of \emph{BuraliForti.v} is where the package shows what makes Rocq immune. The author defines what looks like a witness for the hypothesis:
\begin{lstlisting}
Record A0 : Type :=
  i0 { X0 : Type ; R0 : X0 -> X0 -> Prop }.
\end{lstlisting}

This is the obvious encoding: \texttt{A0} is the type of pairs of a type and a relation on it, and \texttt{i0} is the constructor. The lemma \texttt{inj} is then provable by case analysis on the record. One would naively expect \texttt{Definition Paradox : False := Burali\_Forti A0 i0 inj} to compile and produce a contradiction. It does not. The Rocq kernel raises a Universe Inconsistency error. The reason is documented in the closing comment:

The constructor \texttt{i0} carries a type \texttt{X0 : Type} inside the record, so the record itself lives in a universe strictly above the universe of its members. Call the inner universe \emph{Type\_j'} and the outer universe \emph{Type\_i'}; the kernel imposes \emph{Type\_j' $<$ Type\_i'} because \texttt{A0} admits large elimination on the record. The Burali-Forti construction, on the other hand, needs to instantiate the hypothesis with \texttt{Omega := i0 A0 emb}, which requires \emph{j $\geq$ i}: the universe of the inner type must be at least as large as the universe in which \texttt{A0} itself lives. The two constraints are inconsistent, and the kernel refuses to assemble them. The paradox is, in a precise sense, the price the kernel pays not to admit.

There is a small piece of irony here, and it is worth pausing on. The Rocq kernel does not need to be told that Burali-Forti is impossible. It does not check the paradox and reject it on the grounds that something has gone wrong. It refuses the construction at the level of universe arithmetic, which is to say it refuses for a reason that has nothing to do with the logical content of the paradox itself. The universe constraints are an engineering decision about how to manage size in the type theory, made for reasons of stratification, and the consequence is that Burali-Forti cannot be coded. The boundary is structural rather than ad hoc, and the package's contribution is to make this visible: the comment at the end of \emph{BuraliForti.v} is, as much as anything else, a piece of formal documentation of what the kernel is doing.

%%==============================================================
\section{Diaconescu: Choice Implies Excluded Middle}
\label{sec:diaconescu}

The Diaconescu paradox is shorter than Burali-Forti, sharper in its conclusion, and more disturbing in its consequences for any constructive system that hopes to admit choice. The argument, in the form due to \citet{werner1997sets} that \emph{diaconescu.v} mechanises, takes an apparently mild choice principle and extracts the law of excluded middle from it as a corollary. The mildness of the hypothesis is what makes the result striking: one is not assuming the full set-theoretic axiom of choice, but only a typed version that asks for a choice function on every equivalence relation. The conclusion is the same. The result has antecedents in the original set-theoretic argument of \citet{diaconescu1975axiom} and in the parallel intuitionistic argument of \citet{goodman1978choice}, but Werner's contribution was to make the construction visible inside CIC as a piece of code.

The Werner mechanisation starts by defining what it means for \texttt{R : A -> A -> Prop} to be an equivalence relation:
\begin{lstlisting}
Definition EquivRel (A : Set) (R : A -> A -> Prop) :=
  (forall x : A, R x x) /\
  (forall x y : A, R x y -> R y x) /\
  (forall x y z : A, R x y -> R y z -> R x z).
\end{lstlisting}

The choice principle is then stated as a quantifier over equivalence relations:
\begin{lstlisting}
Definition Tchoice :=
  forall (A : Set) (R : A -> A -> Prop),
    EquivRel A R ->
    ex (fun f : A -> A =>
      (forall x : A, R x (f x)) /\
      (forall x y : A, R x y -> f x = f y)).
\end{lstlisting}

The principle says that for every equivalence relation \texttt{R} on a set \texttt{A}, there exists a function \texttt{f : A -> A} that picks a representative from each equivalence class, in the sense that \texttt{R x (f x)} holds for every \texttt{x}, and that representatives are stable under the relation, in the sense that \texttt{R x y} implies \texttt{f x = f y}. This is the typed version of the axiom of choice: it is asking for a uniform way of selecting representatives, not for a non-uniform choice function in the set-theoretic sense.\footnote{The natural reading of ``reasonable axiom of choice'' in type theory has to be careful. The choice principle that \citet{werner1997sets} refutes is a \emph{typed} choice principle: it produces a function \texttt{f : A -> A} rather than picking elements from sets. The reason this matters is that the function \texttt{f} is by definition extensional, so \texttt{R x y -> f x = f y} follows from the typing rather than from any separate hypothesis. This is what makes the argument carry through in CIC and what distinguishes the Diaconescu paradox from the classical set-theoretic story. The set-theoretic Diaconescu argument needs the axiom of extensionality of sets explicitly; the type-theoretic version gets it for free from the way functions are defined.}

The argument then proceeds by considering the fixed type \texttt{bool} and the parameter proposition \texttt{P : Prop} whose decidability is to be proved. The inductive type \texttt{rel : bool -> bool -> Prop} is defined as follows:
\begin{lstlisting}
Inductive rel : bool -> bool -> Prop :=
  | rrefl : forall b : bool, rel b b
  | rel2  : forall b c : bool, P -> rel b c.
\end{lstlisting}

Two booleans are related either because they are equal (the constructor \texttt{rrefl}) or because the proposition \texttt{P} holds (the constructor \texttt{rel2}). The second constructor is the trick. It collapses every pair of booleans into the same equivalence class whenever \texttt{P} is true, and leaves the relation as the identity when \texttt{P} is false. The lemmas \texttt{rel\_sym}, \texttt{rel\_trans}, and the combined \texttt{rel\_equiv : EquivRel bool rel} are immediate.\footnote{The reader who notices that \texttt{rel2} is a second constructor for \texttt{rel} whose type does not mention the bool indices is observing the trick. The constructor allows any pair of booleans to be related, provided P holds. The resulting inductive type therefore collapses to a single equivalence class when P is true, and to two singleton classes when P is false. The proof of EM is then a question about which of these two situations the choice function \texttt{f} actually lives in, and decidable equality on booleans (which CIC gives without any extra assumption) settles it.}

The application of choice to \texttt{rel} gives a function \texttt{f : bool -> bool} with \texttt{rel x (f x)} for every \texttt{x}, and the extensionality clause \texttt{rel x y -> f x = f y}. The decidable equality of booleans then settles the question. If \texttt{f true = f false}, then by transitivity \texttt{rel true false}, and the inductive cases of \texttt{rel} force \texttt{P} to be inhabited (the \texttt{rrefl} case is impossible because \texttt{true $\neq$ false}, so the \texttt{rel2} case fires). If \texttt{f true $\neq$ f false}, then by the extensionality clause \texttt{$\neg$ (rel true false)}, which means \texttt{$\neg$ P} (because if \texttt{P} held, \texttt{rel true false} would hold by \texttt{rel2}). Either way, \texttt{P $\vee$ $\neg$ P}. The theorem \texttt{EM : P $\vee$ $\neg$ P} is then a closed term parameterised by the choice hypothesis.

The structure of the argument is purely computational. There is no metaphysical claim about choice or about excluded middle being made in the proof; there is a small program that, given a choice function on bool, computes a decision for \texttt{P}. The decision is made by comparing two booleans and reading off the result. The Diaconescu paradox in this presentation is the observation that the program is well-typed, and that its existence as a closed term means the hypothesis \texttt{Tchoice} cannot be added to CIC without making excluded middle derivable.

This has a consequence for how the Rocq kernel handles propositional content. The kernel does admit excluded middle as a hypothesis (it is consistent with CIC), and it does admit a variety of choice principles. What it does not admit is the conjunction of \texttt{Tchoice} with constructive disjunction at the level of \texttt{Set}: that combination would import the law of excluded middle into \texttt{Set}, and the Hurkens paradox in Section~\ref{sec:hurkens} will show that this combination is inconsistent. The Diaconescu file is therefore a piece of the case against admitting choice in \texttt{Set}, and it is striking how small the case is. Twenty lines of code and a decidable equality on booleans is enough to show the price.

The file closes with a second variant, \texttt{TTDiaconescu2}, that uses a slightly different formulation of choice (via existence rather than a function) and proves the same conclusion. The variant is included presumably because the original formulation of the paradox in the set-theoretic literature uses the existence form, and the file wants to be transparent about the equivalence. The substance is the same. The two presentations are connected by the observation that a choice function on equivalence classes is the same datum as a system of representatives, and the latter is what one extracts from existential choice in CIC.

%%==============================================================
\section{Reynolds: Polymorphism Is Not Set-Theoretic}
\label{sec:reynolds}

The Reynolds paradox is the longest file in the package, running to nearly six hundred lines, and the argument it formalises is by some distance the most intricate of the four. The result, in the form due to \citet{reynolds1984polymorphism}, says that there is no set-theoretic model of System F: no way to interpret the polymorphic types of System F as actual sets in a way that respects the operational semantics of the calculus. The file \emph{Reynolds.v} reformulates this as a result about Coq's Prop universe, which is impredicative in the same way System F's type quantifier is impredicative, and proves that no injection from Prop into a single proposition can exist (Barras and Werner, in \citealp{barras2019paradoxes}).

The hypothesis under refutation is the existence of an injection \texttt{I : Prop -> Heyt}, where \texttt{Heyt} is some fixed proposition. The injection is required to satisfy:
\begin{lstlisting}
Hypothesis Iinject :
  forall (P Q : Prop), I P = I Q -> (P <-> Q).
\end{lstlisting}

The reading is that two propositions whose images under \texttt{I} coincide must be logically equivalent. The aim is to derive False from this hypothesis. The structure of the argument is, at a high level, an instance of Cantor's diagonal theorem: there cannot be a surjection from a set \texttt{A} onto its powerset \texttt{A -> Heyt}, and one will be exhibited if the hypothesis \texttt{Iinject} holds. The execution of the argument has to be done with care because the surjection is constructed inside CIC, where the powerset is not directly available and has to be approximated by a complete Heyting algebra.

The first move is to define a retract structure. The injection \texttt{I : Prop -> Heyt} is given by hypothesis; the partial inverse \texttt{T : Heyt -> Prop} is constructed as \texttt{T b := exists P : Prop, P /\textbackslash{} b = I P}. The lemmas \texttt{E1 : x -> T (I x)} and \texttt{E2 : T (I x) -> x} make \texttt{I} and \texttt{T} a bijection between Prop and the image of \texttt{I}, and the rest of the development uses this bijection to transport the structure of Prop's Heyting algebra onto \texttt{Heyt}. The relation \texttt{Heyt\_Eq x y := T x <-> T y} makes \texttt{Heyt} into a setoid, and the file proves that \texttt{Heyt\_Eq} is reflexive, symmetric, and transitive (a per, in the file's terminology, where the per is in fact a full equivalence). The setup is laborious but necessary: one cannot work with quotients in Coq without naming the relevant equivalence, and the file is being scrupulous about which equality is in use at which point.\footnote{The structure \texttt{A0 := forall A : Prop, (PHI A -> A) -> A} is what category theorists call a preinitial algebra for the functor PHI, and what type theorists recognise as the Church encoding of an inductive type. The reason the development needs to refine A0 to a quotient by the partial equivalence relation E0 is that the Church encoding is only \emph{pre}initial: the universal property holds up to provable equality, but not up to definitional equality, and Cantor's diagonal argument needs the latter. The quotient supplies the missing extensionality, and the cost of the quotient is that one has to thread the per through every construction. The development at this point begins to look like ordinary categorical algebra written in Coq syntax.}

The substantive content begins with the definition of the Reynolds functor:
\begin{lstlisting}
Definition PHI (A : Prop) := (A -> Heyt) -> Heyt.
\end{lstlisting}

The functor \texttt{PHI} sends a proposition \texttt{A} to the type of ``subsets of \texttt{A}'' (functions \texttt{A -> Heyt}) composed with the powerset construction (functions from those into \texttt{Heyt}). It is the type-theoretic analogue of the double-powerset functor on sets, and the Cantor argument will be set up against its preinitial algebra. The functorial action on functions is the obvious one:
\begin{lstlisting}
Definition phi {A B : Prop} (f : A -> B) (z : PHI A) : PHI B :=
  fun u : B -> Heyt => z (fun x : A => u (f x)).
\end{lstlisting}

The preinitial algebra for \texttt{PHI} is the Church-encoded type:
\begin{lstlisting}
Definition A0 : Prop := forall A : Prop, (PHI A -> A) -> A.
\end{lstlisting}

This is the type of natural transformations from \texttt{PHI} to \texttt{Id} in Prop, and it is the universal solution to the equation \texttt{X = PHI X} if one is willing to read ``universal'' at the level of provable equivalence. The algebra structure is given by:
\begin{lstlisting}
Definition A0_cons (z : PHI A0) : A0 :=
  fun (A : Prop) (f : PHI A -> A) => f (phi (A0_ind f) z).
\end{lstlisting}

with \texttt{A0\_ind} the induction principle that any element of \texttt{A0} packages by definition. The function \texttt{A0\_match := A0\_ind (phi A0\_cons)} supplies a partial inverse to \texttt{A0\_cons}, in the sense that \texttt{A0\_cons} and \texttt{A0\_match} are inverse up to provable equality on the quotient by the per \texttt{E0} that the file constructs. The fact that an initial algebra of a functor \texttt{F} satisfies \texttt{F X $\cong$ X} up to isomorphism is the content of \citeauthor{lambek1968fixpoint}'s \citeyearpar{lambek1968fixpoint} fixed-point theorem, which is the categorical statement of the property being engineered here.

The \texttt{E0} relation is defined as the smallest per on \texttt{A0} for which \texttt{A0\_cons} is a set-function and \texttt{A0\_match} is its inverse:
\begin{lstlisting}
Definition E0 (x1 x2 : A0) : Prop :=
  forall E : Rel A0,
    per E ->
    (forall z1 z2, phi2 E z1 z2 -> E (A0_cons z1) (A0_cons z2)) ->
    (forall u : A0, E u u -> E u (A0_cons (A0_match u))) ->
    E x1 x2.
\end{lstlisting}

The reader who has seen the impredicative encoding of inductive types in System F will recognise this construction. The relation \texttt{E0} is the intersection of all per's that satisfy the two closure conditions, and the encoding is impredicative in the same way \texttt{A0} itself is. The file proves the consequences in sequence: \texttt{per\_E0} (it is a per), \texttt{A0\_cons\_set\_func} (\texttt{A0\_cons} respects the per), \texttt{id\_A0\_cons\_match} (\texttt{A0\_cons} and \texttt{A0\_match} are inverse on reflexive elements), and finally \texttt{A0\_match\_set\_func} (\texttt{A0\_match} respects the per). The quotient of \texttt{A0} by \texttt{E0} is then an initial \texttt{PHI}-algebra in the category of setoids, which is the structure Reynolds's argument needs.

With the initial algebra in hand, the Cantor argument can begin. The file constructs a surjection \texttt{khi : A0 -> (A0 -> Heyt)} by composing the inverse \texttt{A0\_match : A0 -> PHI A0} with the intersection operation \texttt{intersect : PHI A0 -> (A0 -> Heyt)}, which sends a subset of subsets to its intersection:
\begin{lstlisting}
Definition intersect (C : PHI A0) (x : A0) : Heyt :=
  I (forall P : A0 -> Heyt, F0 P P -> T (C P) -> T (P x)).
\end{lstlisting}

The companion injection \texttt{singleton : (A0 -> Heyt) -> PHI A0} sends a subset to the singleton family containing it, and the lemma \texttt{id\_intersect\_singleton} shows that intersection is a retraction of singleton on reflexive subsets. By composition, \texttt{khi} is a surjection onto the set of reflexive subsets of \texttt{A0}.

The Lawvere fixed-point theorem then closes the argument. The theorem says that any function \texttt{f : Heyt -> Heyt} has a fixed point relative to a surjection from \texttt{A0} onto \texttt{A0 -> Heyt}:
\begin{lstlisting}
Lemma Lawvere_fixpoint : forall f : Heyt -> Heyt,
  let q := (fun a : A0 => f (khi a a)) in
  F0 q q ->
  let p := psi q in
  Heyt_Eq (f (khi p p)) (khi p p).
\end{lstlisting}

Applying the theorem to the function \texttt{f h := I ($\neg$ T h)} (the negation in the Heyting algebra structure on \texttt{Heyt}) produces a fixed point of negation. But negation has no fixed points (the lemma \texttt{negnofix} in the file proves this), so False follows \citep{lawvere1969diagonal}. The theorem \texttt{Reynolds : False} is closed in fewer than ten lines once the infrastructure is in place.

There is a corollary that deserves to be mentioned, because it converts the Reynolds paradox into a statement about proof irrelevance. The lemma \texttt{EM\_PI} at the end of \emph{Reynolds.v} says that excluded middle implies proof irrelevance: if \texttt{forall P : Prop, P $\vee$ $\neg$ P}, then for every proposition \texttt{Q} and any two proofs \texttt{a, b : Q}, one has \texttt{a = b}. The proof uses excluded middle to define an injection from Prop into \texttt{Q} by the formula \texttt{I P := if EM P then a else b}; if \texttt{a $\neq$ b}, this injection satisfies the Reynolds hypothesis, and so produces False, and from False one derives \texttt{a = b}. The corollary is, in some sense, the most striking consequence of the package: classical logic in Coq's Prop forces proof irrelevance, which is a property the kernel of Rocq does not impose by default and which has substantial consequences for the way classical mathematics is formalised in the system.\footnote{\citeauthor{reynolds1984polymorphism}'s \citeyearpar{reynolds1984polymorphism} original argument was in the setting of System F semantics, and what he showed was that no set-theoretic functor \emph{T} can have a fixed point in the strong sense required by polymorphic types. The Coq reformulation by Barras and Werner restates the impossibility in terms of an injection \texttt{I : Prop -> Heyt} into a single proposition; the underlying combinatorial content is the same. The advantage of the Coq version is that it places the paradox inside a proof assistant that is, by construction, immune to it. The kernel rejects the hypothesis \texttt{Iinject} on universe grounds before any of the rest can be evaluated, which is what allows the file to exist in the library as a piece of working mathematics rather than as a contradiction.}

%%==============================================================
\section{Hurkens in Impredicative Set: The Cost of Classical Logic}
\label{sec:hurkens}

The last of the four files in the package, \emph{Hurkens\_Set.v}, mechanises an adaptation by \citet{geuvers2001inconsistency} of \citeauthor{hurkens1995simplification}'s \citeyearpar{hurkens1995simplification} simplification of Girard's paradox. The original Hurkens paradox is a derivation of False in system U-minus, and its remarkable feature is its compactness: in fewer than thirty lines of typed lambda calculus, Hurkens exhibits a closed term inhabiting the empty type. Geuvers's adaptation transports this argument into the setting of the Calculus of Inductive Constructions with impredicative Set, and shows that excluded middle at Set is inconsistent with the rest of CIC.

The argument hinges on the construction of a retract between Prop and a small type. A retract, in the technical sense the file uses, is a pair of maps \texttt{p2b : Prop -> bool} and \texttt{b2p : bool -> Prop} together with two compatibility hypotheses. The first hypothesis says that \texttt{p2b} embeds Prop into bool up to double negation:
\begin{lstlisting}
Hypothesis p2p1 : forall A : Prop, dn (b2p (p2b A)) -> dn A.
\end{lstlisting}

where \texttt{dn A := (A -> False) -> False} is the double-negation modality. The second hypothesis is the converse direction:
\begin{lstlisting}
Hypothesis p2p2 : forall A : Prop, A -> b2p (p2b A).
\end{lstlisting}

The two together make \texttt{p2b} a kind of partial inverse to \texttt{b2p} up to double negation, which is enough for the Hurkens construction to fire. The construction itself is a remarkable piece of compressed code:
\begin{lstlisting}
Definition V := forall A : Set, ((A -> bool) -> A -> bool) -> A -> bool.
Definition U := V -> bool.
Definition sb (z : V) : V := fun A r a => r (z A r) a.
Definition le (i : U -> bool) (x : U) : bool :=
  x (fun A r a => i (fun v => sb v A r a)).
Definition induct (i : U -> bool) : Prop :=
  forall x : U, b2p (le i x) -> dn (b2p (i x)).
Definition WF : U := fun z => p2b (induct (z U le)).
\end{lstlisting}

The reader who tries to parse these definitions by their stated types will find that they are deliberately compressed. The type \texttt{V} is a polymorphic type that quantifies over a set \texttt{A}, a predicate transformer \texttt{(A -> bool) -> A -> bool}, and an element \texttt{a : A}, returning a boolean. The type \texttt{U} is a function from \texttt{V} to booleans. The function \texttt{sb} is a substitution operation on \texttt{V}; the function \texttt{le} is an order relation on \texttt{U} encoded as a boolean-valued function. The predicate \texttt{induct} expresses well-foundedness with respect to \texttt{le}, and \texttt{WF : U} is the well-founded element constructed from this predicate. The Hurkens construction is a self-referential definition: \texttt{WF} is well-founded with respect to a relation that itself is defined in terms of \texttt{WF}, and the impredicativity of Set is exactly what allows this self-reference to be coded.\footnote{\citeauthor{hurkens1995simplification}'s \citeyearpar{hurkens1995simplification} original paradox is a remarkable piece of compression: a derivation of False in system U-minus in less than thirty lines of typed lambda calculus. \citeauthor{geuvers2001inconsistency}'s \citeyearpar{geuvers2001inconsistency} adaptation transports the argument to the setting of CIC with impredicative Set and excluded middle, and the file \emph{Hurkens\_Set.v} in the coq-paradoxes library is the mechanisation of Geuvers's argument. The mechanisation is more verbose than Hurkens's original (the file runs to 163 lines), but the verbosity is largely bookkeeping for the double-negation translation that Geuvers uses to extract the contradiction from the retract. The structure of the argument is unchanged from 1995.}

The contradiction is reached through three lemmas. The lemma \texttt{Omega} says that every well-founded predicate \texttt{i} satisfies \texttt{dn (b2p (i WF))}; the lemma \texttt{lemma} says that the specific predicate \texttt{I} (defined locally in the proof) satisfies \texttt{induct}; the lemma \texttt{lemma2} says that \texttt{(forall i, induct i -> dn (b2p (i WF))) -> False}. Composing these gives the theorem \texttt{Hurkens\_set\_neg : False} under the retract hypotheses.

The application to excluded middle is direct. Suppose, for contradiction, that \texttt{EM\_set\_neg : forall A : Prop, \{$\neg$ A\} + \{$\neg\neg$ A\}} holds in CIC. Define \texttt{p2b A := if EM\_set\_neg A then false else true} and \texttt{b2p b := b = true}. The hypotheses \texttt{p2p1} and \texttt{p2p2} are then provable by case analysis on \texttt{EM\_set\_neg A}, and the Hurkens construction fires to produce False. The file's intermediate theorem \texttt{not\_EM\_set\_neg} records this, and a final section then notes that the strictly stronger \texttt{\{A\} + \{$\neg$ A\}} trivially entails the weaker negation form, and so it too is inconsistent:
\begin{lstlisting}
Section EM_set_inconsistency.
  Variable EM_set_neg : forall A : Prop, {A} + {~ A}.
  Theorem not_EM_set : False.
End EM_set_inconsistency.
\end{lstlisting}

That is, excluded middle in Set is inconsistent with the rest of CIC. The kernel of Rocq does not derive False from excluded middle in Set, of course, because the kernel does not admit \texttt{forall A, \{A\} + \{$\neg$ A\}} as a hypothesis; what the file shows is that admitting it would be a mistake. The boundary the kernel draws is again precise: classical disjunction is admissible in Prop (where it merely fails to be constructive), but not in Set (where it would import a small-type witness of any classical fact, and where Hurkens shows that the witnesses break the system).

There is a structural reason this happens at Set rather than at Prop, and it deserves to be stated. The universe Set in CIC is impredicative in the same way Prop is: a quantifier \texttt{forall A : Set, ...} that ranges over the entire universe is allowed to live in Set itself. The difference between Prop and Set is that Set permits large elimination (one can define a function from a Set into Type), and Prop does not. The combination of impredicativity with large elimination is exactly what Hurkens needs: the self-referential definition of \texttt{WF} requires both the impredicative quantification (to define \texttt{V} and \texttt{U}) and the large elimination (to extract booleans from propositional content via \texttt{p2b}). The Rocq kernel admits impredicativity at Prop only, and admits large elimination at Set only, and the two are kept apart precisely so that the Hurkens construction cannot be assembled. The file is a piece of evidence that this design decision has been made for the right reasons.

%%==============================================================
\section{The Three Boundary Conditions}
\label{sec:boundary}

The four files of the coq-paradoxes package can be read separately, as four pieces of mechanised mathematics that happen to derive False under appropriate hypotheses. They can also be read together, as a single document that draws three boundary conditions on the Calculus of Inductive Constructions \citep{coquand1988calculus, coquand1990inductively}. The article has been moving toward this second reading, and the remaining work is to state the conditions explicitly.\footnote{The dependency between the four files is worth registering. \emph{Logics.v} provides the well-foundedness machinery used by \emph{BuraliForti.v}. \emph{Log\_Rel.v} provides the partial-equivalence-relation machinery used by \emph{Reynolds.v}. The files \emph{diaconescu.v} and \emph{Hurkens\_Set.v} are largely self-contained, depending only on the standard library. The shared infrastructure is therefore concentrated in well-foundedness and partial equivalences, both of which are standard tools in any non-trivial CIC formalisation. The package is not so much a collection of paradoxes as a collection of failed encodings, each of which would have been a paradox in a less careful system.}

The first boundary condition is the universe hierarchy. The Burali-Forti paradox in \emph{BuraliForti.v} is, in its essential content, a result about what happens when one allows a type to quantify over itself in a relational way. The hypothesis \texttt{i0 : forall X : Type, (X -> X -> Prop) -> A0} is innocent in isolation, but combined with the wish to take \texttt{Omega := i0 A0 emb} it forces \texttt{A0} to live in a universe that contains its own image, which is a fixed-point property of universes that classical type theories refuse. The Rocq kernel implements the refusal by stratifying its universes and by recording cumulativity constraints during type checking; the constraint \emph{Type\_j' $<$ Type\_i'} that the kernel raises when one tries to encode \texttt{A0} is the mechanical implementation of this refusal. The boundary is, in a precise sense, the price of having a fixed-point operator on universes \citep{girard1972interpretation}.

The second boundary condition is the placement of impredicativity. The Reynolds paradox in \emph{Reynolds.v} and the Hurkens paradox in \emph{Hurkens\_Set.v} both turn on the question of where in the universe hierarchy impredicative quantification is allowed. Prop is impredicative in CIC, and the impredicativity is what allows the Church-encoded preinitial algebra \texttt{A0 := forall A : Prop, (PHI A -> A) -> A} in Reynolds and the polymorphic types \texttt{V} and \texttt{U} in Hurkens. Set, on the other hand, is impredicative only in CIC's impredicative-Set extension, and the Hurkens paradox shows that adding impredicativity to Set is harmless only as long as one does not also admit excluded middle in Set. The default Rocq kernel keeps Set predicative, which avoids the issue entirely; the impredicative-Set extension is available as a flag, and the package is a piece of documentation of what the flag costs.

The third boundary condition is the discipline of large elimination. The Diaconescu paradox in \emph{diaconescu.v} and the Hurkens paradox both use large elimination at the boundary between propositional and computational content. In Diaconescu, the choice function \texttt{f : bool -> bool} is extracted from an existence proof, which is a form of large elimination: a function in Set is produced from a witness in Prop. In Hurkens, the retract function \texttt{p2b : Prop -> bool} is the same kind of move, and it is exactly the move that imports excluded middle in Set into the system. The Rocq kernel allows large elimination only out of certain inductive types and out of \texttt{Set}, and never out of \texttt{Prop} unless the inductive type in Prop has at most one constructor (so that the elimination is trivially propositional). The boundary is again precise: the kernel knows which large eliminations are safe and refuses the rest.

These three boundary conditions are not independent. The Reynolds paradox depends on impredicativity in Prop, which is also what makes the Hurkens paradox possible at Set when impredicativity is admitted there. The Burali-Forti paradox depends on the universe hierarchy, but the same hierarchy is what allows the kernel to refuse the unsafe large eliminations that would otherwise let Diaconescu run. The three conditions sit together as a single coordinated policy on the kernel, and the four paradoxes are four ways of running into the policy from different directions.

There is a methodological observation worth stating in closing. The coq-paradoxes package is a piece of mathematics that is, by design, never executed as a proof of False. The package compiles, and the theorems it proves go through, but each theorem is parameterised by a hypothesis the kernel rejects. The hypotheses are stated as \texttt{Variable} or \texttt{Hypothesis} declarations inside sections, which means the kernel never has to commit to them; the closing comments in each file show what would happen if one tried to. This is, on inspection, an unusual idiom for a mathematics package. Most Coq libraries assume their hypotheses and prove their theorems; the paradoxes package assumes hypotheses that cannot be satisfied and proves theorems whose only role is to document the unsatisfiability. The package is, in a precise sense, a piece of metamathematics written inside the system whose metatheory it documents, and it is one of the more elegant pieces of documentation in the Rocq ecosystem \citep{rocqdev2025}.

What the four paradoxes share, when read together, is a particular relationship between the user and the kernel. The user writes a construction that would, in a less careful system, produce False. The kernel notices the construction at the level of universe arithmetic, large elimination, or impredicativity, and refuses it. The refusal is not a special case; it is a consequence of policies that the kernel applies uniformly to all input. The four paradoxes are four occasions on which the policies do what they were designed to do, and the package is the formal record of the design having worked. This is not a small achievement, and it is not obvious from outside that it has been achieved; the paradoxes package, taken together, is the evidence.

%%==============================================================
\bibliographystyle{plainnat}
\bibliography{references}

@misc{barras2019paradoxes,
  author = {Barras, Bruno and Coquand, Thierry and Herbelin, Hugo and Werner, Benjamin},
  title  = {\texttt{coq-paradoxes} 8.10.0},
  year   = {2019},
  note   = {Coq library, Rocq Prover Package Index},
  url    = {https://rocq-prover.org/p/coq-paradoxes/latest}
}

@inproceedings{coquand1986analysis,
  author    = {Coquand, Thierry},
  title     = {An analysis of {G}irard's paradox},
  booktitle = {Proceedings of the First Annual {IEEE} Symposium on Logic in Computer Science ({LICS} 1986)},
  pages     = {227--236},
  year      = {1986},
  publisher = {IEEE Computer Society Press},
  address   = {Washington, DC}
}

@article{coquand1988calculus,
  author  = {Coquand, Thierry and Huet, G{\'e}rard},
  title   = {The calculus of constructions},
  journal = {Information and Computation},
  volume  = {76},
  number  = {2--3},
  pages   = {95--120},
  year    = {1988},
  doi     = {10.1016/0890-5401(88)90005-3}
}

@incollection{coquand1990inductively,
  author    = {Coquand, Thierry and Paulin-Mohring, Christine},
  title     = {Inductively defined types},
  booktitle = {{COLOG-88}},
  editor    = {Martin-L{\"o}f, Per and Mints, Grigori},
  series    = {Lecture Notes in Computer Science},
  volume    = {417},
  pages     = {50--66},
  year      = {1990},
  publisher = {Springer},
  address   = {Berlin},
  doi       = {10.1007/3-540-52335-9_47}
}

@article{diaconescu1975axiom,
  author  = {Diaconescu, Radu},
  title   = {Axiom of choice and complementation},
  journal = {Proceedings of the American Mathematical Society},
  volume  = {51},
  number  = {1},
  pages   = {176--178},
  year    = {1975},
  doi     = {10.2307/2040105}
}

@misc{geuvers2001inconsistency,
  author = {Geuvers, Herman},
  title  = {Inconsistency of classical logic in type theory},
  year   = {2001},
  note   = {Note, Radboud University Nijmegen},
  url    = {https://www.cs.ru.nl/%7Eherman/PUBS/newnote.pdf}
}

@phdthesis{girard1972interpretation,
  author = {Girard, Jean-Yves},
  title  = {Interpr{\'e}tation fonctionnelle et {\'e}limination des coupures de l'arithm{\'e}tique d'ordre sup{\'e}rieur},
  school = {Universit{\'e} Paris VII},
  year   = {1972}
}

@article{goodman1978choice,
  author  = {Goodman, Noah D. and Myhill, John},
  title   = {Choice implies excluded middle},
  journal = {Zeitschrift f{\"u}r mathematische Logik und Grundlagen der Mathematik},
  volume  = {24},
  number  = {25--30},
  pages   = {461},
  year    = {1978},
  doi     = {10.1002/malq.19780242514}
}

@inproceedings{hurkens1995simplification,
  author    = {Hurkens, Antonius J. C.},
  title     = {A simplification of {G}irard's paradox},
  booktitle = {Typed Lambda Calculi and Applications ({TLCA} 1995)},
  editor    = {Dezani-Ciancaglini, Mariangiola and Plotkin, Gordon},
  series    = {Lecture Notes in Computer Science},
  volume    = {902},
  pages     = {266--278},
  year      = {1995},
  publisher = {Springer},
  address   = {Berlin},
  doi       = {10.1007/BFb0014058}
}

@article{lambek1968fixpoint,
  author  = {Lambek, Joachim},
  title   = {A fixpoint theorem for complete categories},
  journal = {Mathematische Zeitschrift},
  volume  = {103},
  pages   = {151--161},
  year    = {1968},
  doi     = {10.1007/BF01110627}
}

@incollection{lawvere1969diagonal,
  author    = {Lawvere, F. William},
  title     = {Diagonal arguments and {C}artesian closed categories},
  booktitle = {Category Theory, Homology Theory and their Applications {II}},
  series    = {Lecture Notes in Mathematics},
  volume    = {92},
  pages     = {134--145},
  year      = {1969},
  publisher = {Springer},
  address   = {Berlin},
  doi       = {10.1007/BFb0080769}
}

@techreport{reynolds1984polymorphism,
  author      = {Reynolds, John C.},
  title       = {Polymorphism is not set-theoretic},
  number      = {RR-0296},
  year        = {1984},
  institution = {INRIA},
  type        = {Research Report},
  url         = {https://hal.inria.fr/inria-00076261}
}

@inproceedings{werner1997sets,
  author    = {Werner, Benjamin},
  title     = {Sets in types, types in sets},
  booktitle = {Theoretical Aspects of Computer Software ({TACS} 1997)},
  editor    = {Abadi, Mart{\'\i}n and Ito, Takayasu},
  series    = {Lecture Notes in Computer Science},
  volume    = {1281},
  pages     = {530--546},
  year      = {1997},
  publisher = {Springer},
  address   = {Berlin},
  doi       = {10.1007/BFb0014566}
}

@misc{rocqdev2025,
  author = {{Rocq Development Team}},
  title  = {The {R}ocq {P}rover Reference Manual, Version 9.1.0},
  year   = {2025},
  note   = {INRIA},
  url    = {https://rocq-prover.org}
}

\end{document}